\documentclass[aps,prd,amsmath,showpacs,twocolumn,nofootinbib]{revtex4}

\newcommand{\be}{\begin{eqnarray}}
\newcommand{\ee}{\end{eqnarray}}
\newcommand{\nn}{\nonumber\\}

\usepackage{graphicx}

\usepackage{subfigure}

\begin{document}
\title{Pion masses in quasiconformal gauge field
theories}
\author{Dennis D.~Dietrich}
\author{Matti J\"arvinen}
\affiliation{Center for High Energy Physics, University of
Southern Denmark, Odense, Denmark}
\date{\today}
\begin{abstract}
We study modifications to Weinberg-like sum rules in quasiconformal gauge
field theories. Beyond the two Weinberg sum rules and the oblique $S$
parameter we study the pion mass and the $X$ parameter. Especially, we 
evaluate the pion mass for walking technicolour theories, in particular 
also minimal walking technicolour, and find contributions of the order of 
up to several hundred GeV.
\end{abstract}
\maketitle


{\it Introduction.}
The Large Hadron Collider (LHC) has been built to help clarifying the
mechanism for the breaking of the electroweak symmetry. Beyond the standard
model, where the symmetry is broken by an elementary scalar Higgs particle,
technicolour (TC) \cite{TC} provides a possible mechanism which overcomes some of the
deficiencies of the former. In TC the electroweak symmetry is
broken by chiral symmetry breaking among additional fermions charged under
the electroweak and the TC gauge groups. Three of the possibly
more numerous Nambu--Goldstone modes are absorbed as longitudinal degrees of
freedom of the electroweak gauge bosons. Walking \cite{WTC}, that
is, quasiconformal TC theories with techniquarks in higher
dimensional representations are compatible with currently available
precision data \cite{Dietrich:2005jn,Dietrich:2006cm}. Like everywhere in strongly
interacting gauge theories effective Lagrangians are heavily used
\cite{EFFECT,Foadi:2007ue}. In many cases the predictive power of these approaches is
enhanced by enforcing the Weinberg sum rules \cite{EFFECT,Foadi:2007ue,Weinberg:1967kj}.

Asymptotically free gauge field theories obey the Weinberg sum rules 
\cite{Weinberg:1967kj}. It has been shown, however, that the second sum rule
is modified \cite{Appelquist:1998xf} in quasiconformal theories while the
expression for the first remains unchanged. In the same sense also the
oblique $S$ parameter \cite{Peskin:1990zt} remains unchanged. Once the
modified second sum rule is imposed together with the unmodified first,
though, the $S$ parameter is reduced in quasiconformal theories
\cite{Appelquist:1998xf}. Here we study related quantities, concretely, the
$X$ parameter \cite{Barbieri:2004qk} and the electroweak contributions to the pion masses \cite{Das:1967it,Peskin:1980gc}. The bare expression for the $X$ parameter does not
receive any corrections, but its value is reduced after the modified second 
sum rule is enforced simultaneously. To the contrary, the 
pion mass is directly modified in quasiconformal theories. 

In the context of TC said mass is that of the
Nambu--Goldstone modes which are not absorbed as longitudinal degrees of
freedom of the electroweak gauge bosons. First of all, the sign of the
squared mass decides 
whether the embedding of the electroweak gauge group in the flavour symmetry
group is stabilised or destabilised by electroweak radiative corrections
\cite{Peskin:1980gc}. In the latter case an additional 
mechanism is needed to stabilise the theory and make it complete. Secondly,
to date, no degrees of freedom which could correspond to these modes have
been detected. Therefore, they have to be sufficiently massive to have
elluded detection. Hence, a sizeable contribution from the electroweak
sector is phenomenologically advantageous. Below we will see that quasiconformal 
dynamics lead indeed to an enhancement of the aformentioned magnitude.


~\\

{\it Pion mass.}
The Weinberg sum rules are given by \cite{Weinberg:1967kj},
\be
\int\frac{ds}{s}~\mathrm{Im}~\Pi(s)=0
\mathrm{~~~and~~~}
\int ds~\mathrm{Im}~\Pi(s)=0,
\ee
respectively, where
\be
\Pi^{ab}_{\mu\nu}(q)
=
\left(g_{\mu\nu}-\frac{q_\mu q_\nu}{q^2}\right)\delta^{ab}\Pi(q^2),
\ee
and
\be
\Pi^{ab}_{\mu\nu}
=
-i\int d^4x e^{-iq\cdot x}
\langle V^a_\mu(x)V^b_\nu(0)-A^a_\mu(x)A^b_\nu(0)\rangle.
\ee
$V^a_\mu$ and $A^a_\mu$ stand for the vector and axial-vector currents,
respectively, with flavour index $a\in\{1,\dots,N_f^2-1\}$. 
 The oblique $S$ 
parameter \cite{Peskin:1990zt} reads,
\be
\frac{S}{4\pi}=\int\frac{ds}{s^2}~\mathrm{Im}~\bar{\Pi}(s),
\ee
where $\bar{\Pi}$ is $\Pi$ without the contribution from Goldstone bosons.
The $X$ parameter \cite{Barbieri:2004qk} is obtained from
\be
\frac{X}{4\pi f_\pi^2}=\int\frac{ds}{s^3}~\mathrm{Im}~\bar{\Pi}(s).
\ee

The pion mass matrix falls into two factors \cite{Peskin:1980gc} 
\be
m^2_{ab}=\hat{m}^2_{ab}\times m^2,
\ee
where $\hat{m}^2_{ab}$ depends on the embedding of the electroweak
gauge group in the flavour symmetry group and $m^2$ stands for the
overall magnitude \cite{Das:1967it}, 
\be
m^2
=-\frac{3\left(g_1^2+g_2^{2}\right)}{64 \pi^2 f_\pi^2} \int_0^\infty \! ds \log s\ \mathrm{Im}\bar{\Pi}(s)
,
\ee
where $g_1$ and $g_2$ are the weak coupling constants and $f_\pi$ is the pion decay constant. 
For the last relation the result for the second Weinberg sum rules has been
used and is important to guarantee the convergence of the integral. 
The overall magnitude $m^2$ will be influenced by the quasiconformal dynamics. 

The chiral symmetry breaking is to proceed from the unbroken flavour
symmetry group $\mathcal{G}$ to the residual group $\mathcal{H}$. The
generators spanning $\mathcal{H}$ are to be called $S^a$ while the rest of
the generators of $\mathcal{G}$ be called $X^a$. 
Following the definition of \cite{Peskin:1980gc} we normalize  
the generators $\Lambda^i$ to which the electroweak fields couple as
\be
 {\cal L} = \sum_i \bar \psi \gamma^\mu  \Lambda^i\psi\, A^i_\mu ,
\ee
where 
$\psi$ is the techniquark field, $\Lambda^i$ is a matrix in flavour space and the sum is over the electroweak gauge fields.
Notice that the $\Lambda^i$ also include the corresponding coupling constants.
They are linear combinations of both categories of generators, 
$\Lambda^i=\Lambda^i_S+\Lambda^i_X$, where 
$\Lambda^i_S$ stands for the contribution from unbroken generators and $\Lambda^i_X$ for 
the contribution from broken generators. Then 
\be \label{massfact}
\hat{m}^2_{ab}
&=& \frac{8}{g_1^2+g_2^2} \sum_i {\rm Tr}\left\{\left[\Lambda^i_S,\left[\Lambda^i_S,X^a\right]\right]X^b\right. - \nn
&& - \left.\left[\Lambda^i_X,\left[\Lambda^i_X,X^a\right]\right]X^b \right\} ,
\ee
where we used the normalisation $ {\rm Tr} X^aX^b = \delta^{ab}/2 =  {\rm Tr} S^aS^b$.

For a running theory and assuming saturation by the lightest resonances, we
have
\be
\mathrm{Im}~\Pi_\mathrm{run}(s)
&=&
\mathrm{Im}~\bar{\Pi}_\mathrm{run}(s)-f_\pi^2s\delta(s),\nn
\mathrm{Im}~\bar{\Pi}_\mathrm{run}(s)&=&F_V^2\delta(s-M_V^2)-F_A^2\delta(s-M_A^2),
\ee
where $F_V$ and $F_A$ are the vector and axial-vector decay constants,
respectively. Multiple resonances can be included straightforwardly by
summing over them. For this spectral function we obtain from the sum rules,
\be
\frac{F_V^2}{M_V^2}-\frac{F_A^2}{M_A^2}=f_\pi^2
\mathrm{~~~and~~~}
F_V^2-F_A^2=0,
\ee
for the oblique parameters
\be
\frac{F_V^2}{M_V^4}-\frac{F_A^2}{M_A^4}&=&\frac{S}{4\pi},\\
\frac{F_V^2}{M_V^6}-\frac{F_A^2}{M_A^6}&=&\frac{X}{4\pi f_\pi^2},
\ee
and for the common factor of the pion masses,
\be
m^2_\mathrm{run}
=
\frac{3\left(g_1^2+g_2^{2}\right)}{64 \pi^2 f_\pi^2}
\frac{F_V^2}{f_\pi^2}\ln\frac{M_A^2}{M_V^2}.
\ee

The picture laid down in \cite{Appelquist:1998xf} for the spectral function
in walking theories is the following: For scales below the continuum
threshold which for $s$ is $O[4\Sigma(0)^2]$, where $\Sigma(p)$ stands for the
dynamical mass of the fermions, the contributions are coming from distinct
resonances like in $\Pi_\mathrm{run}(s)$. From this scale up to $\Lambda^2$,
where the system drifts away from conformality again, a continuum of quarks
and gluons governs the form of the spectral function. Let us denote this
part as $\Delta(s)$. Contributions beyond $\Lambda^2$ are negligible, such
that
\be
\mathrm{Im}~\Pi_\mathrm{walk}=\mathrm{Im}~\Pi_\mathrm{run}(s)+\Delta(s).
\ee
It has to be noted, that the the first addend above is supposed to have a
structure like in the running case, but need not be identical down to the
values of the parameters.
It turns out that in the first sum rule,
\be
\int \frac{ds} {s} \Delta(s)\approx 0
\ee
because the integral is too concentrated around the origin in order to
be sensitive to the modification. For the same reason, the expression for the $S$
and $X$ parameters are even less affected. To the contrary \cite{Appelquist:1998xf},
\be
\int ds~\Delta(s)=-8 \pi^2 \alpha f_\pi^4/d_\mathrm{R},
\ee
where $d_\mathrm{R}$ is the dimension of the representation of the gauge
group under which the fermions transform. $\alpha$ is expected to be
positive and of order one. The contribution to $m^2$ is given by,
\be
m^2_\mathrm{walk}
=
\frac{3\left(g_1^2+g_2^{2}\right)}{64 \pi^2 f_\pi^2}
\left[
F_V^2\ln\frac{\lambda^2}{M_V^2}
-
F_A^2\ln\frac{\lambda^2}{M_A^2}
\right].
\ee
The scale $\lambda$ is defined through the condition
\be
\int_{4\Sigma^2(0)}^{\Lambda^2}ds\ln\frac{\lambda^2}{s}\Delta(s)=0.
\label{scale}
\ee
Hence, $\lambda\in[2\Sigma(0),\Lambda]$ and it can be estimated to lie arround the geometric mean of $2\Sigma(0)$ and $\Lambda$.

Next, we use the first and the modified second Weinberg sum rule to
eliminate the decay constants from the expressions for $S$, $X$ 
and the pion mass:
\be
\frac{S}{4\pi}
=
f_\pi^2\frac{M_V^2+M_A^2}{M_V^2M_A^2}
-
\alpha\frac{8\pi^2f_\pi^4}{d_\mathrm{R}}\frac{1}{M_V^2M_A^2},\nonumber
\ee
\be
\frac{X}{4\pi f_\pi^2}
=
f_\pi^2\frac{M_V^4+M_V^2M_A^2+M_A^4}{M_V^4M_A^4}
-
\alpha\frac{8\pi^2f_\pi^4}{d_\mathrm{R}}\frac{M_V^2+M_A^2}{M_V^4M_A^4},
\nonumber
\ee
\be
m^2_\mathrm{walk}
=
\frac{3\left(g_1^2+g_2^{2}\right)}{64 \pi^2}
\left[
\frac{M_V^2M_A^2}{M_A^2-M_V^2}\ln\frac{M_A^2}{M_V^2}
\right.
+
\nn
+
\alpha\frac{8\pi^2}{d_\mathrm{R}}\frac{f_\pi^2}{M_A^2-M_V^2}
\left(
M_A^2\ln\frac{\lambda^2}{M_A^2}
\right.
-
\left.\left.
M_V^2\ln\frac{\lambda^2}{M_V^2}
\right)
\right].
\nonumber
\ee
For given values of $M_V$ and $M_A$, the $S$ \cite{Appelquist:1998xf} and 
$X$ parameters are reduced in walking theories with respect to running
theories (with $\alpha=0$). As long as $M_V$ and $M_A$ are smaller than $\mathrm{e}\lambda^2$,
which due to $M_{A/V}^2<4\Sigma(0)^2<\lambda^2$ (see the more detailed 
explanation above) can be assumed, the pion mass is enhanced for 
walking theories. 


~\\

{\it Walking technicolour.}
Let us study what this means for TC theories. For these
$f_\pi^2=2\Lambda_\mathrm{ew}^2/N_f^g$ where $\Lambda_\mathrm{ew}=246$GeV.
$N_f^g$ stands for the number of techniflavours gauged under the electroweak
interactions. This number need not be equal to the number of
techniflavours. In a situation where more than two techniquarks are needed
to achieve walking dynamics it is advantageous to gauge only two of them
under the electroweak gauge group in order to alleviate the pressure from
the constraints on the oblique $S$ parameter. This setup is known as
partially gauged TC \cite{Dietrich:2005jn}.

\begin{figure*}[t]
\subfigure{\includegraphics[width=4.4cm]{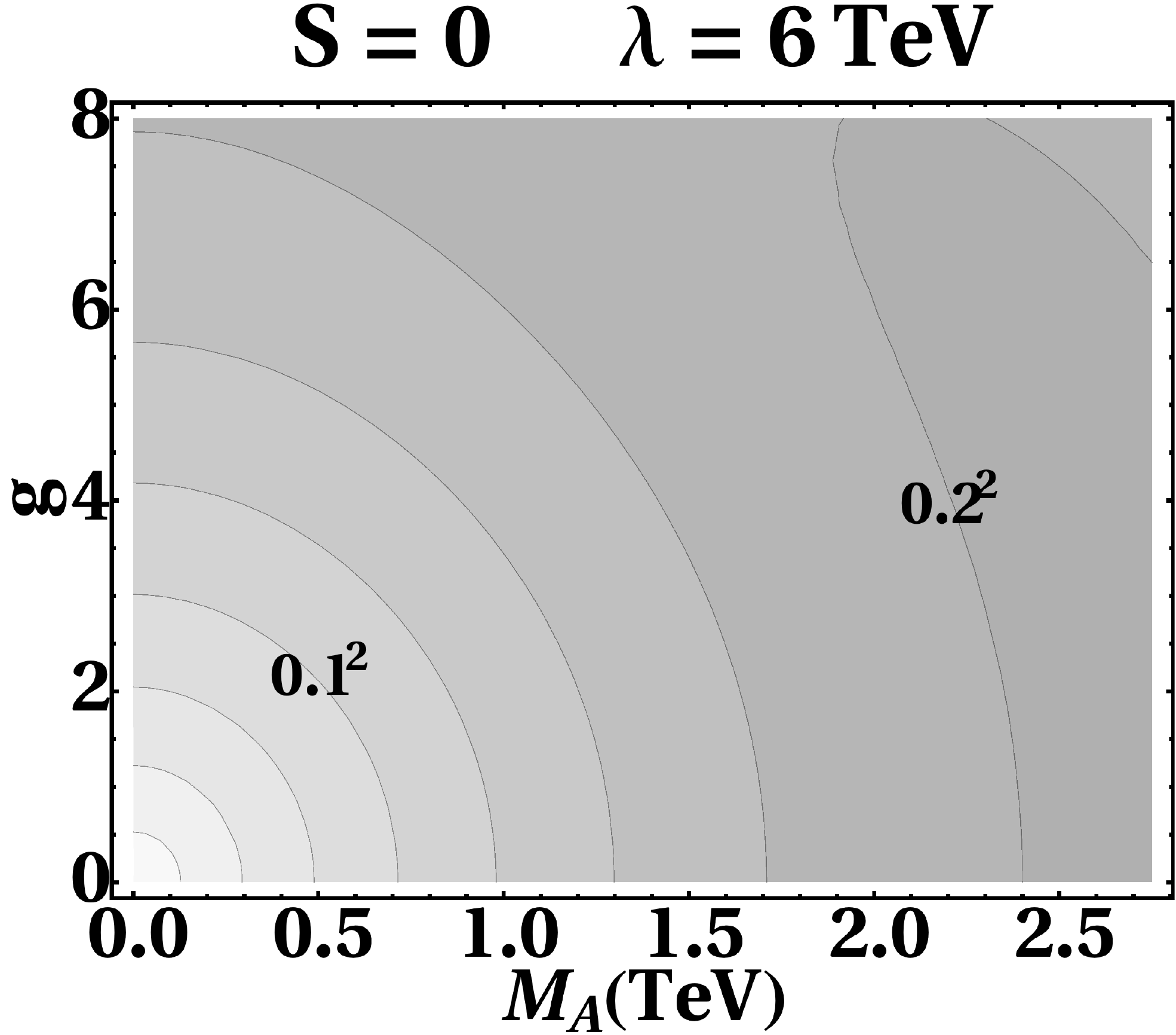}}
\subfigure{\includegraphics[width=4.4cm]{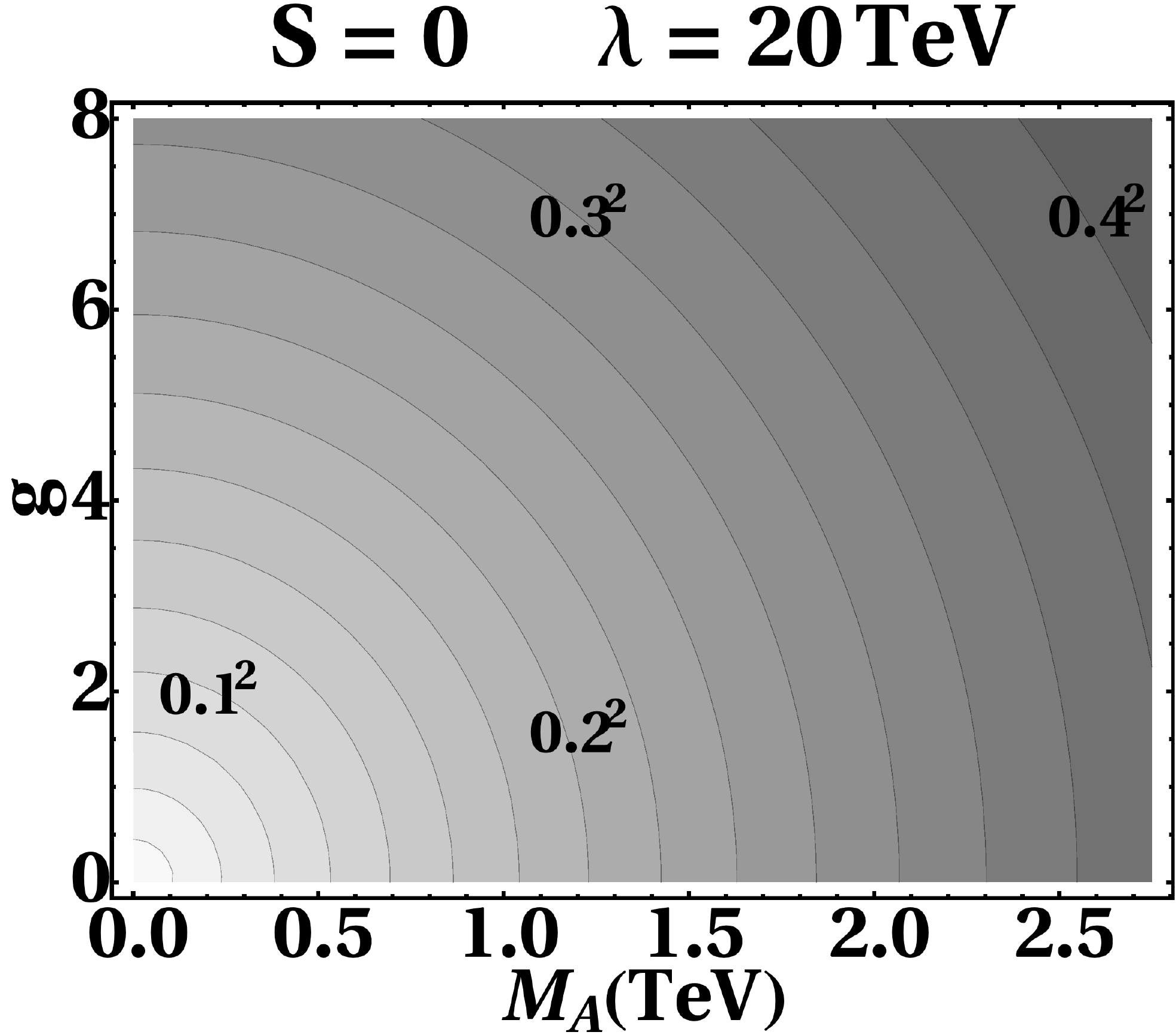}}
\subfigure{\includegraphics[width=4.4cm]{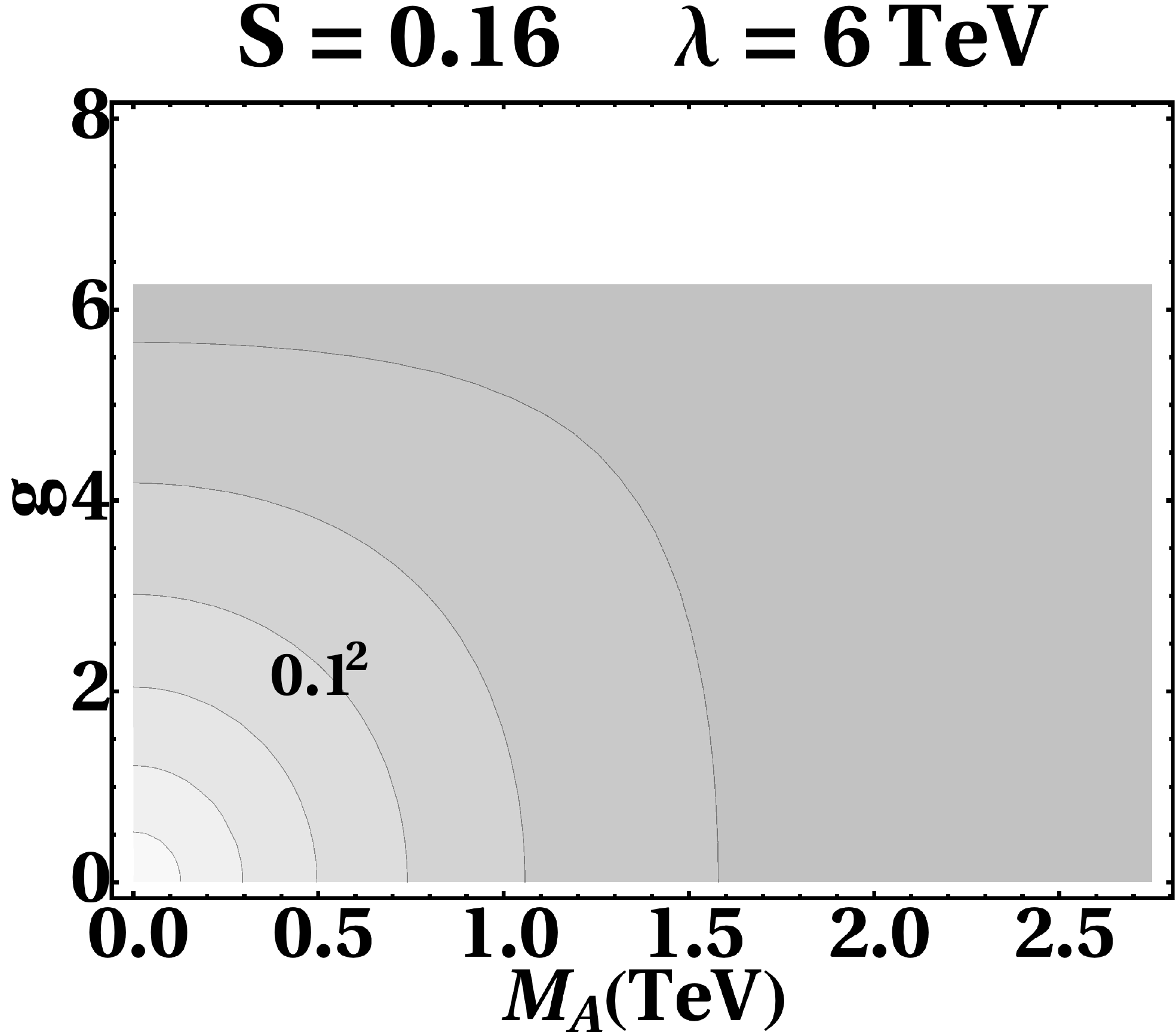}}
\subfigure{\includegraphics[width=4.4cm]{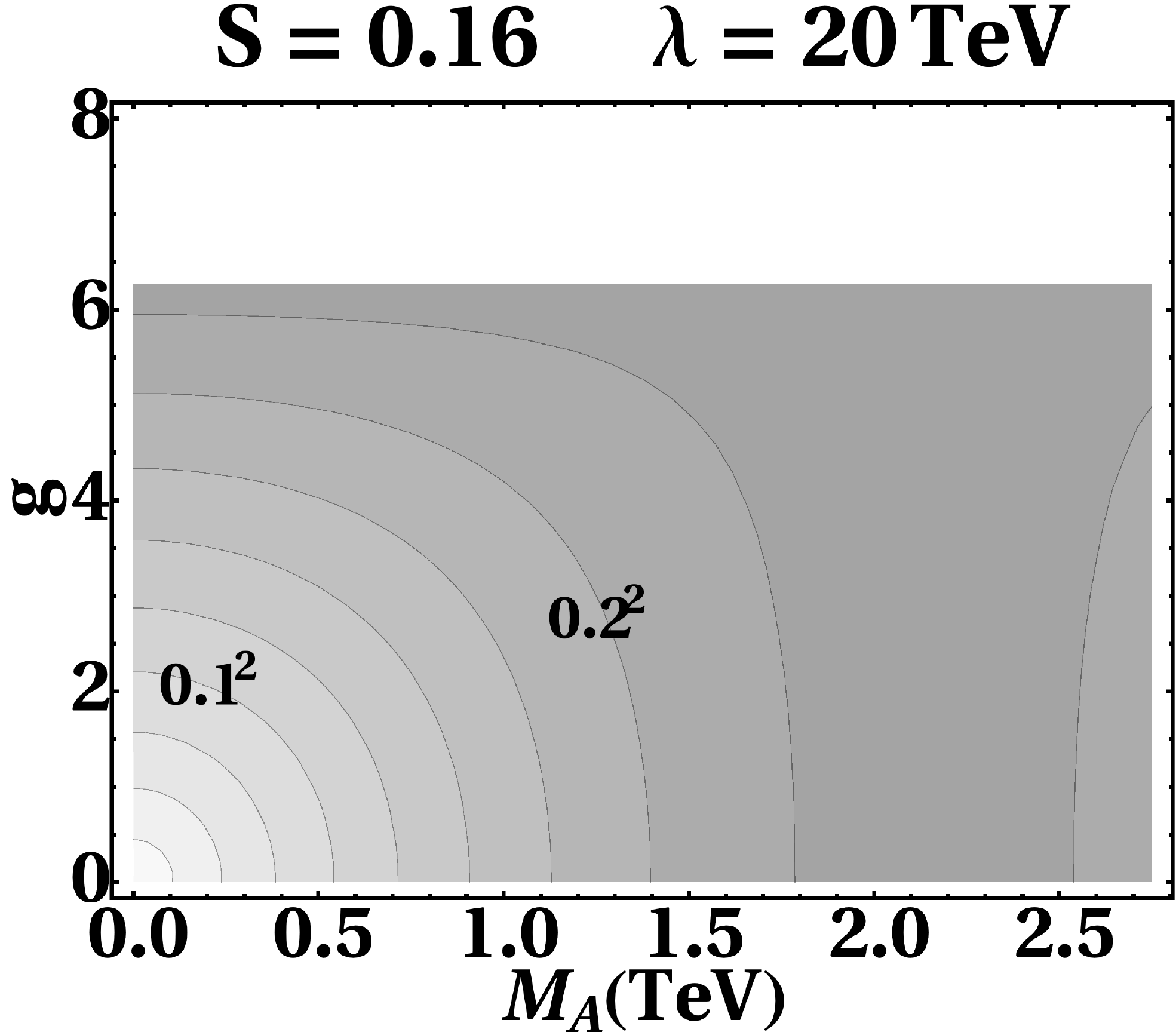}}
\caption{
Contour plots of $m^2_\mathrm{walk}$ as function of $M_A$ and $g=M_V^2/(\sqrt{2}F_V)$ for $S=0$ ($S=S_\mathrm{pert}^\mathrm{MWT}\approx0.16$) and $\lambda=6$TeV ($\lambda=20$ TeV). 
Contours are for $m^2_\mathrm{walk} = 0.025^2$~TeV$^2$, $0.05^2$~TeV$^2$,  $0.075^2$~TeV$^2$, \dots with zero at the origin.
For $S=S_\mathrm{pert}^\mathrm{MWT}$, the plot range in $g$ is constrained by the requirement $F_A^2>0$.
}
\label{Fig}
\end{figure*}

Apart from imposing the first Weinberg sum rule we introduce a coupling $g$
in the spirit of the coupling connected to a local symmetry such that 
$M_V^4=2g^2F_V^2$. It is expected to be of order unity. 
For positive $S$ there is an upper limit for $g$ above which $F_A^2$ turns negative.
In the plots, we display $m^2_\mathrm{walk}$ which is a model 
independent factor. The detailed mass structure $\hat m^2_{ab}$ differs
between models. 
The normalisation of $m^2_\mathrm{walk}$ has been chosen such 
that the eigenvalues of $\hat m^2_{ab}$ are in general of order unity.
Hence, one can directly gain an initial impression on the magnitude of the pion
masses from the numbers in the plots in Fig.~\ref{Fig}: For
small $S$ parameters and large values of $\lambda$, $M_A$ and $g$ up to
400GeV can be reached, which corresponds to five times the mass of the W
boson. 

For a concrete model, MWT
features two
techniquarks transforming under the adjoint representation of an $SU(2)$
gauge group. The adjoint representation is real. The unbroken chiral
symmetry is therefore enhanced to $SU(4)$ which breaks to $SO(4)$ yielding
nine Goldstone modes, three of which become the longitudinal degrees of
freedom of the weak gauge bosons. This leaves behind six extra technipions
which here carry nonzero baryon number. In this setup, the techniquarks
alone would lead to a theory with a global Witten anomaly, as an odd number
of fermions [The adjoint of $SU(2)$ is three-dimensional.] would transform
under the $SU(2)_L$ of the Standard Model. The anomaly is cancelled by
including an additional pair of leptons. Due to the presence of these
leptons the requirement of freedom from gauge anomalies does not fix the
hypercharge assignment completely, but leaves one free continuous parameter.

By using the generators of MWT from \cite{Dietrich:2005jn,Foadi:2007ue} in (\ref{massfact}) we find for
the electroweak contributions to the masses of the technipions of MWT 
which are not eaten by the electroweak gauge fields,
\be
 \left(g_1^2+g_2^2\right) \hat m^2_{\Pi_{UU} } &=& g_1^2\left(1+2y\right)^2 +g_2^2 \nn 
\left(g_1^2+g_2^2\right) \hat m^2_{\Pi_{DD} } &=& g_1^2\left(1-2y\right)^2 +g_2^2 \nn 
\left(g_1^2+g_2^2\right) \hat m^2_{\Pi_{UD} } &=& g_1^2\left(4y^2-1\right) +g_2^2 .
\ee
Here $y$ is a parameter that controls the hypercharge assignment of the techniquarks and the subscripts $U$ and $D$ denote the flavours of the techniquarks which constitute the corresponding technipion.
The masses of the charge conjugate pions are the same. 
Since $g_2>g_1$ the squared masses are positive for all values of $y$.
In Table~\ref{table:masses} we present the numerical values for the most natural choices of $y$. 
$g_1$ and $g_2$ are weakly dependent on the mixing of the electroweak gauge bosons with the composite vector states in MWT. We use their standard model (tree-level) values which is 
sufficient for our purposes. 

\begin{table}
\begin{eqnarray}
\begin{array}{c||ccc}
 y &\quad \hat m^2_{\Pi_{UU} }  &\quad \hat m^2_{\Pi_{DD} }  &\quad \hat m^2_{\Pi_{UD} } \\ 
\hline\hline
0 & \quad  1 &\quad 1  &\quad 0.53 \\
+1 & \quad  2.87 &\quad 1  &\quad 1.47 \\
-1 & \quad  1 &\quad 2.87  &\quad 1.47 
\end{array} \nonumber
\end{eqnarray}
\caption{Eigenvalues of $\hat m_{ab}^2$ for the uneaten technipions in MWT for various values of the parameter $y$. }
\label{table:masses}
\end{table}

The squares of the physical masses as functions of $M_A$ and $g$ are obtained by multiplying the values 
in Fig.~\ref{Fig} by the factors of Table~\ref{table:masses}. Pion masses 
for the light mass window \cite{Foadi:2007ue} of MWT, where $M_A<M_V \lesssim 1$~TeV, are from 50 to 300 GeV. 
For the heavy mass window, where $M_A>M_V \gtrsim 2$~TeV, all pion masses are at least 150 GeV. 
Interestingly, the pion masses can be large enough to exceed present experimental bounds even without 
any additional extended technicolour (ETC) interactions, expect in the region of small $M_A$ and $g$.

The low-energy limit of the continuum is given by twice the dynamical
mass $\Sigma(0)$, i.e., the threshold of the loop. It's estimated to be
$2\Sigma(0)\approx 4\pi f_\pi/d_\mathrm{R}$. For MWT
 this evaluates to circa 1 TeV. Picking resonances considerably
above this value entails a slight modification of the picture, if one does
not want to put them inside the continuum: The continuum threshold has to be
taken higher. The only definite constraint for our investigation is that the
scale $\lambda$ must lie inside the continuum. As we have chosen reference
values of $\lambda=6$ TeV and $\lambda =20$ TeV, we extend our plots to
close to 3 TeV resonance masses.

The model known under the name Next-to-Minimal Walking Technicolour (NMWT)
possesses two techniquarks which transform under the two-index symmetric
representation of SU(3) which is not (pseudo)real, but even-dimensional.
Consequently, the flavour symmetry breaking pattern is $SU(2)_L\times
SU(2)_R\rightarrow SU(2)_V$ leading to three Goldstone modes which are all
eaten by the weak gauge bosons. Hence, there are no degrees of freedom left
which would receive contributions to their mass in the framework of the
present analysis.

Other viable candidates for dynamical electroweak symmetry breaking
by walking TC theories are discussed in \cite{Dietrich:2006cm} and
can be analysed using the above formulae.


~\\

{\it Summary.}
We studied the modifications of the Weinberg sum rules in quasiconformal gauge field theories. 
We showed that while the $X$ parameter is not directly altered by nearly conformal dynamics, 
the electroweak corrections to pion masses can be enhanced considerably wrt. running theories. 
As an explicit example we discussed MWT where the typical scale of 
these corrections was seen to be from 100 to 200~GeV.
Interestingly, the technipion masses can be large enough to exceed present experimental bounds even without any additional ETC interactions, expect in the region of small $M_A$ and $g$.

The naive expectation that the masses (not the squared masses) at least in
the running case can be simply scaled up from their value in quantum
chromodynamics to their value in techincolour by the ration say of the
respective pion decay constants is not confirmed. For the QCD pions the
above formulation returns the difference if the squared masses of the
charged and the neutral pions. In order to obtain the linear mass
difference, the result has to be divided by the sum of the masses, which is
dominated by explicit symmetry breaking, i.e., the bare quark masses.
In the application to TC directly the squared masses of the
uneaten pions are computed, the square root of which returns the linear
masses. Therefore, it is understandable that the thus obtained pion masses
may (and do) exceed the values obtained by scaling up the QCD masses.
Quasiconformal dynamics leads to an additional enhancement.


~\\

{\it Acknowledgments.}
The authors acknowledge useful discussions with
Roshan Foadi,
Mads T.~Frandsen,
Chris Kouvaris,
and
Francesco Sannino.
The work of DDD was supported by the Danish Natural Science Research
Council.
The work of MJ was supported by the Marie Curie Excellence Grant under
contract MEXT-CT-2004-013510.


\end{document}